\documentclass[9pt,twocolumn,twoside]{osajnl}

\journal{ol} 

\setboolean{shortarticle}{true}

\title{Measuring optical vortices by means of dual shearing-type Sagnac interferometers}

\author[1]{Han-Tao Wang}
\author[1]{Hua-Jun Zhang}
\author[1]{Ming-Yuan Ren}
\author[1]{Wen-kai Yao}
\author[1,*]{Yu Zhang}

\affil[1]{School of Physics, Harbin Institute of Technology, No.92, Xidazhi Road, Harbin 150001, China}

\affil[*]{zhangyuhitphy@163.com}

\begin{abstract}
Measuring the positions of optical vortices is an essential part in the researches of speckles and adaptive optics. The measurement accuracy is restricted by the performance of optical devices and the properties of optical vortices, such as density and size. In order to achieve high accuracy and wide range of application, the dual shearing-type Sagnac interferometers is proposed using two shearing plates to adjust the precision of optical vortices measurement. The shearing displacements are able to balance the measuring precision and the value of the intensity ratio point to provide optimum measurement performance. This method is useful for the observation of optical vortices with different sizes and densities, especially for the high density condition.
\end{abstract}

\setboolean{displaycopyright}{true}

\begin{document}

\maketitle

Since the concept of optical vortices was introduced into optics by Nye and Berry \cite{nye1974dislocations}, the properties of optical vortices have been intensively studied in adaptive optics \cite{1998Branch}, speckle patterns \cite{2008Fractality} and astrophysics \cite{2008Method}. Choosing appropriate methods for measuring the topological charges and the locations of optical vortices is a crucial part of experimental research. Due to the instability of high-order charged optical vortex \cite{1999Critical}, only the optical vortices with $\pm1$ topological charges are widespread in the random interference field of coherent light besides rare doubly degenerate vortices \cite{1993Optical}. That shifts the focus of measurement to the distribution of optical vortices. Measuring the points with zero magnitude directly is insufficient for optical vortices detection because of the limitations of the distinguishability of low intensity region \cite{Baranova1983Wave}. Reference wave based interferometric method with high spatial resolution is a simple way for wave front measurement \cite{2011Examination}. However, due to the non-common path structure, it is susceptible to vibrations and results in the reduction of accuracy. Thus, as a noninterferometric method, Shack-Hartmann wave front sensor (SHWFS) overcomes the problems and decrease the complexity of optical systems \cite{2015Detection}. Several optical vortex detection methods were proposed for use with a SHWFS, such as vortex potential method, contour sum method and the slope discrepancy technique \cite{2012Comparison}. But its unequalled performance for optical vortices detection is limited to the subaperture size \cite{Mingzhou2008Dipole}. For example, closer optical vortices need smaller subaperture size for detection. And that requires more pixels of a complementary metal oxide semiconductor (CMOS) sensor to achieve higher accuracy \cite{wu2019wish}.

In order to achieve flexible and high accurate measurement, a common path optical system with adjustable measuring precision is needed. For adjustable precision, it has been demonstrated that multi-pinhole interferometer is suitable for measuring optical vortices of arbitrary sizes \cite{2008Method,2010Measuring}. But the precondition requires the optical vortex to be in the center of the holes. That is unrealistic for optical vortices with random distributions. In fact, the essence of multi-pinhole interferometer is analyzing the wave front slopes around the optical vortex through interference patterns. This common-path multi-wave interference can be replaced by several shearing interderometers connected in series \cite{2005wave,2010Common}. To further achieve precision adjustment, in this paper, we proposed a method to measure optical vortices based on dual shearing-type Sagnac interferometers using two adjustable shearing plates.

\begin{figure}[ht!]
\centering
\includegraphics[width=\linewidth]{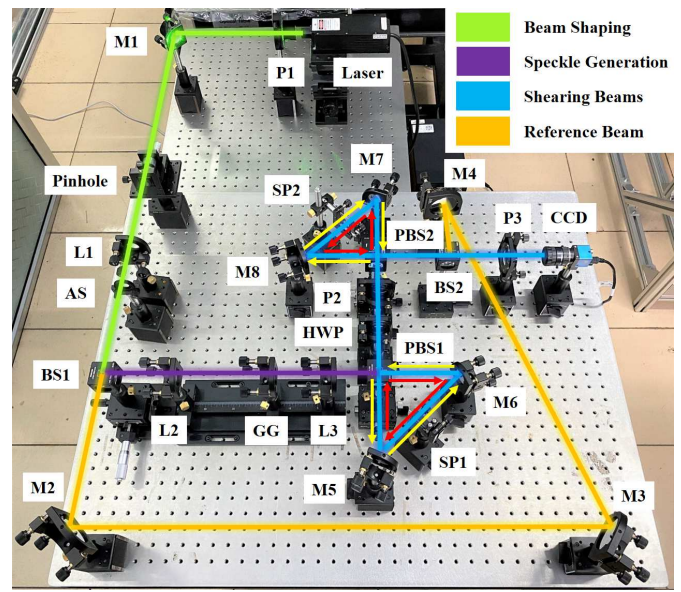}
\caption{Experimental setup of the complete optical systems containing dual shearing-type Sagnac interferometers and a reference wave based interferometer.}
\label{setup}
\end{figure}

The experimental setup is illustrated in Fig. \ref{setup}. The whole optical system includes four parts, namely, beam shaping, speckle generation, dual shearing-type Sagnac interferometers and reference wave based interferometer. In the beam shaping part, the linearly polarized beam generated by a $532\rm{nm}$ single longitudinal mode laser is collimated and expanded to a proper Gaussian beam by the mirror M1, a pinhole filter, the convex lens L1 and the aperture stop AS. And the polarizer P1 is used for the polarization adjustment of the emergent light. Then a $50:50$ beam splitter BS1 divides the Gaussian beam into reference arm and speckle generation arm.

The speckle generation part uses a ground glass plate GG to generate speckles with different speckle sizes through changing the focusing spot size on the GG \cite{goodman2007speckle}. Namely, the density of optical vortices is adjustable due to the variable position of the GG in the 4f system composed of lenses L2 and L3. Then the speckle propagates into the dual shearing-type Sagnac interferometers. Each Sagnac interferometer consists of a polarizing beam splitter PBS, two mirrors and a (Thorlabs, BCP43R) shearing plate SP. Because of the PBS1 and PBS2, the horizontally polarized component of the speckle propagates in the clockwise direction (red arrows). And the vertically polarized component propagates anticlockwise (yellow arrows). For shearing setups, the SP1 in the first Sagnac interferometer is used for horizontal shearing and the SP2 in the second Sagnac interferometer is used for vertical shearing. Between these two Sagnac interferometers, we put a half wave plate HWP for phase adjustment and the P2 for combining two orthogonal polarized beams into a whole. Similarly, the P3 offers the same functionality as the P2. Eventually the interefence pattern of four replicas of the speckle is imaged to a (Daheng Optics, R-125-30UM) monochrome charge-coupled device CCD with $964\times1292$ pixels. 

In order to examine the accuracy of optical vortices measurement, a reference wave based interferometer is set up using the Gaussian beam reflected by M2, M3 and M4 as the reference beam. Removing all the shearing plates and the HWP, the outgoing wave of the shearing beam part can be regarded as a replica of the generated speckle. Coupling the reference beam into the interference path using the BS2 and the P3, then the interference pattern is recorded by the CCD. 

\begin{figure}[ht!]
\centering
\includegraphics[width=0.8\linewidth]{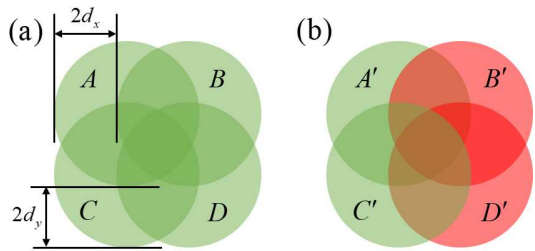}
\caption{Schematic of the combining forms of four replicas.}
\label{scheme}
\end{figure}

To present the interferometric process clearly, a schematic of the combining forms of four replicas is illustrated in Fig. \ref{scheme}. For convenience, the $x$ axis is chosen to be along the horizontal direction and the $y$ axis along the vertical direction. The incoming speckle at PBS1 is simplified into a circular pattern. Due to the opposite directions of two orthogonal polarized beams, the shearing displacements of these two beams have the same length but with opposite directions. Therefore, the horizontal displacement with the length of $d_x$ caused by the SP1 results in the separation of two beams being $2d_x$, for example, the separation of patterns $A$ and $B$ in Fig. \ref{scheme}(a). The SP2 has the same impacts on the divided beams as well but in the vertical directions, respectively. And the separation of the beams with the length of $2d_y$ is shown in Fig. \ref{scheme}(a). 

When $d_x$ and $d_y$ are much smaller than the speckle size, a tiny area in the overlapping region of four replicas can be considered to consist of four small parts of the original speckle in the neighborhood. If the wave front is continuous, we assume that the phase difference of two points in the tiny area can be expressed as
\begin{equation}
\psi \left( {{x_1},{y_1}} \right) - \psi \left( {{x_2},{y_2}} \right) \approx {s_x}\left( {{x_1} - {x_2}} \right) + {s_y}\left( {{y_1} - {y_2}} \right),
\label{eq1}
\end{equation}
where $s_x$ and $s_y$ are the phase slops in the x direction and y direction, respectively. In addition, due to the size of the tiny area, we use the average amplitudes to substitute the amplitudes of the points related to each replica in this area. Namely, $E_A$, $E_B$, $E_C$ and $E_D$ are applied to refer to the amplitudes of four replicas in this tiny area, respectively. If the additional phase shift caused by the HWP only affects one of two orthogonal polarized beams, i.e., the fast axis of the HWP being $0^{\circ}$ or $90^{\circ}$ with respect to the $x$ axis, the intensity of the interference pattern in the presence of the HWP is different from that in the absence of the HWP. For example, in Fig. \ref{scheme}(b), the HWP adds a phase shift on the replica $B'$ (red-filled circle). So does it on the replica $D'$ due to the effect of the SP2. Besides, for optimum interference, the polarization directions of the P1, P2 and P3 are set to be $45^{\circ}$ with respect to the $x$ axis. Based on these preconditions, the intensity difference between the condition with the HWP and that without the HWP can be derived as
\begin{equation}
\begin{split}
\Delta I_{con} =&  - \frac{1}{4}{t}\left[\cos \left( {2{s_x}{d_x}} \right)\left( {E_{A} {E_{B}} + E_{C}{E_{D}}} \right)\right. \\
&+ \cos \left( {2{s_x}{d_x}} \right)\cos \left( {2{s_y}{d_y}} \right)\left( {E_{A } {E_{D}} + E_{B }{E_{C }}} \right)\\
&\left. + \sin \left( {2{s_x}{d_x}} \right)\sin \left( {2{s_y}{d_y}} \right)\left( {E_{A} {E_{D}} - E_{B} {E_{C }}} \right)\right],  
\end{split}
\label{eq2}
\end{equation}
where $t$ is the transmission function of the dual shearing-type Sagnac interferometers. 

Considering an optical vortex existing in this tiny area, the wave front can be regarded as the combination of the continuous phase and the "hidden phase" \cite{Fried:98}. Namely, the wave front consists of the continuous phase as mentioned above and the spiral phase of an optical vortex \cite{1998Branch}. For simplification, the spiral phase is substituted by the phase loop with the distribution of $0$, $\pi/2$, $\pi$ and $3\pi/2$ around the center of the tiny area. And the direction of increasing phase determines the sign of the optical vortex. Note that the interference intensity is associated with the phase difference of two beams rather than the absolute phases of them. Therefore, the intensity difference between the condition with the HWP and that without the HWP is not affected by the attitude of the optical vortex. And it can be written as
\begin{equation}
\begin{split}
\Delta I_{dis} = & - \frac{1}{4}{t}\left[\mp \sin \left( {2{s_x}{d_x}} \right)\left( {E_{A} {E_{B}} - E_{C}{E_{D}}} \right)\right. \\
&- \cos \left( {2{s_x}{d_x}} \right)\cos \left( {2{s_y}{d_y}} \right)\left( {E_{A } {E_{D}} + E_{B }{E_{C }}} \right)\\
&\left. - \sin \left( {2{s_x}{d_x}} \right)\sin \left( {2{s_y}{d_y}} \right)\left( {E_{A} {E_{D}} - E_{B} {E_{C }}} \right)\right],  
\end{split}
\label{eq3}
\end{equation}
where $\mp$ refers to the sign of optical vortex. Using the slowly varying amplitude approximation in this tiny area, for further simplification, Eq. (\ref{eq2}) transforms into
\begin{equation}
\Delta I_{con} =  - \frac{1}{2}{t}E^2\left[\cos \left( {2{s_x}{d_x}} \right) + \cos \left( {2{s_x}{d_x}} \right)\cos \left( {2{s_y}{d_y}} \right)\right],  
\label{eq4}
\end{equation}
where $E$ is the substitution of the average amplitudes of each replica in this area. Similarly, Eq. (\ref{eq3}) transforms into
\begin{equation}
\Delta I_{dis} =  \frac{1}{2}{t}E^2\cos \left( {2{s_x}{d_x}} \right)\cos \left( {2{s_y}{d_y}} \right). 
\label{eq5}
\end{equation}
Comparing Eq. (\ref{eq4}) with Eq. (\ref{eq5}), the intensity variations of these two conditions are completely opposite. And the sign of the optical vortex is no longer the fact to influence the measurement. Besides, all the variations of the intensities are only related to the phase slops of the continuous wave front components. The smaller $s_x$ and $s_y$ lead to more apparent difference between the values of Eq. (\ref{eq4}) and that of Eq. (\ref{eq5}). That is similar to the errors of optical vortex measurement due to the high local tilt of the wave front \cite{Sanchez:11}. However, the phase slop is the nature of the wave front to be measured. Thus, the only way to improve the measuring performance is decreasing the separation of four replicas through adjusting the positions of the shearing plates.

\begin{figure}[ht!]
\centering
\includegraphics[width=0.95\linewidth]{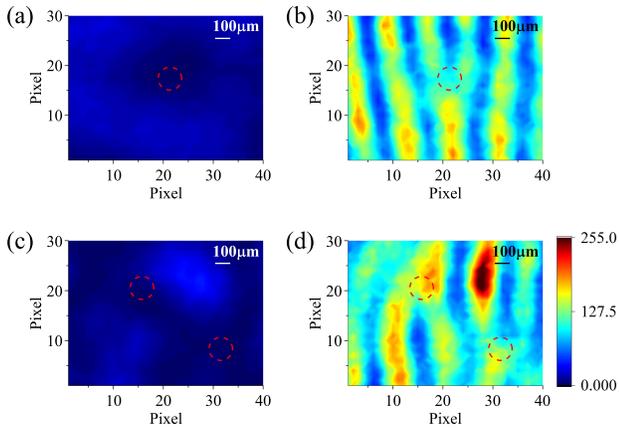}
\caption{(a) The speckle with low optical vortex density and (b) its hologram in a small area. (c) The speckle with high optical vortex density and (d) its hologram in a small area.}
\label{fig3}
\end{figure}

To examine the theoretical conjecture mentioned above, we use the optical system to measure the speckles containing optical vortices with different densities. The setups of the polarizers and the HWP are in accordance with these in the theoretical part. In the absence of shearing plates, two speckles with different speckle sizes are captured. And the interference patterns of the speckles and the reference beam (i.e. hologram) are imaged to measure the locations of optical vortices. In order to present the measuring performance in detail, we choose two small areas as examples, where the optical vortices are clearly visible. These measuring results are shown in Fig. \ref{fig3}.

The speckle with low optical vortex density and its hologram are illustrated in Fig. \ref{fig3}(a) and \ref{fig3}(b). And the speckle with high optical vortex density and its hologram are shown in Fig. \ref{fig3}(c) and \ref{fig3}(d). Although the pixel size of the CCD is $3.75 \mu m \times 3.75\mu m $, for further comparison with $d_x$ and $d_y$, we use the scale bar for the speckles rather than that for the pixel. The optical vortices in two speckles are marked by red dashed circles for the examination of measuring accuracy later. 

We set three groups of $d_x$ and $d_y$ for each speckle measurement. The dimensions of the shearing plate is $25 \rm{mm} \times 36\rm{mm}\times 1\rm{mm}$ and the index of refraction at $532\rm{nm}$ is $n_1=1.46$. For the low optical vortex density group, the angles between the shearing plates and the transverse planes of the transmitted beams are set to be $9.1^{\circ}$ in the horizontal (SP1), $18.2^{\circ}$ in the vertical (SP2) for small shearing displacement, and $16.2^{\circ}$ in the horizontal (SP1), $26.9^{\circ}$ in the vertical (SP2) for large shearing displacement, respectively. For the high optical vortex density group, The angles of large shearing displacement are the same as the low density group. The angles of small shearing displacement are set to be $10.9^{\circ}$ in the horizontal (SP1), $19.4^{\circ}$ in the vertical (SP2). The shearing displacements of thus settings can be calculated by
\begin{equation}
d_j = D\sin \left( \theta_j  \right)\left[ {1 - \frac{{\cos \left( \theta_j  \right)}}{{\sqrt {n_1^2 - {{\sin }^2}\left( \theta_j  \right)} }}} \right], j \in \left\{ {x,y} \right\}, 
\label{eq6}
\end{equation}
where $D$ is the thickness of the shearing plate and $\theta_j$ is the angle mentioned above. Although the displacements caused by shearing plates is able to calculated, the unknown and inevitable displacement resulting from the optical system exists in each experimental groups. Therefore, the measuring results should be analyzed through comparison to eliminate the impact of it. 

\begin{figure}[ht!]
\centering
\includegraphics[width=\linewidth]{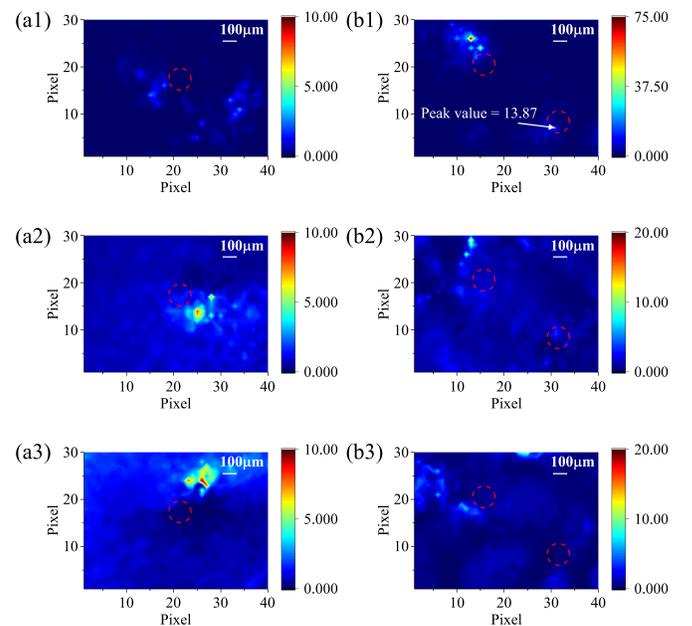}
\caption{The intensity ratio distributions of the conditions with the HWP to those without the HWP.}
\label{fig4}
\end{figure}

We present the measuring results of two speckles with different optical vortex density in Fig. \ref{fig4}. The images are the intensity ratio distributions of the conditions with the HWP to those without the HWP. The intensity ratio distributions of the speckle with low optical vortex density for no shearing plate condition, small shearing displacements and large shearing displacements are illustrated in Fig. \ref{fig4}(a1), \ref{fig4}(a2) and \ref{fig4}(a3), respectively. Similarly, those of the speckle with high optical vortex density are shown in Fig. \ref{fig4}(b1)-\ref{fig4}(b3). The chosen regions are the same as those in Fig. \ref{fig3} for two speckles, respectively. Besides, the locations of the optical vortices are marked in Fig. \ref{fig4} by red dashed circles as well.

According to Eq. (\ref{eq4}) and Eq. (\ref{eq5}), the optical vortex leads to the raise of interference intensity in the neighborhood, but the continuous wave front component has the opposite effect. Therefore, the regions containing optical vortices show higher intensity ratios. That is obvious in Fig. \ref{fig4}(a2) and \ref{fig4}(a3). The separation distances for Fig. \ref{fig4}(a2) are $100\mu m$ in the horizontal and $208\mu m$ in the vertical. And those for Fig. \ref{fig4}(a3) are $184\mu m$ in the horizontal and $324\mu m$ in the vertical. Because the separation distance determines the size of the tiny area in Fig. \ref{scheme}, the larger separation distance means lower accuracy, i.e., the distance between the optical vortex (center of a red dashed circle) and the high intensity ratio point being larger. This phenomenon is also presented in Fig. \ref{fig4}(a2) and \ref{fig4}(a3). Besides, the difference between the deviation of the distance in Fig. \ref{fig4}(a2) and that in Fig. \ref{fig4}(a3) is comparable to the differences of $2d_x$ and $2d_y$ between the condition of small shearing displacements and that of large shearing displacements. That demonstrates the relation between the measuring accuracy and shearing displacements as well from another perspective. However, the intensity ratio distribution in Fig. \ref{fig4}(a1) seems to violate the principle. In fact, the disappearance of the high intensity ratio point is caused by the low intensities around the optical vortex. Comparing the bright areas in Fig. \ref{fig4}(a1) and those in Fig. \ref{fig3}(a), the bright parts in Fig. \ref{fig4}(a1) are the areas around the dark hole in Fig. \ref{fig3}(a). This phenomenon results from the combined effects of the intensities which are not low enough, and the edges of the spiral wave front of the optical vortex. Hence, the optimum $d_x$ and $d_y$ for measuring optical vortices in the speckle with low optical vortex density approach to the states in Fig. \ref{fig4}(a2).

For the measuring results in Fig. \ref{fig4}(b1)-\ref{fig4}(b3), the measuring performance is significantly different from that of the low optical vortex density condition. The separation distances for Fig. \ref{fig4}(b2) are $122\mu m$ in the horizontal and $224\mu m$ in the vertical. And the separation distances for Fig. \ref{fig4}(b3) are the same as those for Fig. \ref{fig4}(a3). The high intensity ratio points are conspicuous in Fig. \ref{fig4}(b1). And the values far larger than those of the low optical vortex density condition. The formation of the points is caused by the inevitable displacement resulting from the optical system. And the high values result from the high gradients of the intensity of the speckle around the optical vortex. Besides, the distance between the optical vortex and the high intensity ratio point is smaller than that in Fig. \ref{fig4}(a2). And it demonstrates the mentioned relation between the measuring accuracy and the shearing displacement. In sharp contrast to this phenomenon, the high intensity ratio point at the upper left corner in Fig. \ref{fig4}(b1) moves away from the optical vortex, and reaches to the location closed to the top edge of Fig. \ref{fig4}(b2). Then, it disappears in Fig. \ref{fig4}(b3). Another high intensity ratio point in Fig. \ref{fig4}(b1) is invisible in Fig. \ref{fig4}(b2) and \ref{fig4}(b3). Besides, the ratios of these two high intensity ratio points decrease gradually with the points fading away. Compared to the low optical vortex density condition, these results demonstrate not only that the shearing displacement influences the measuring accuracy, but also that the smaller the speckle size is, the smaller optimum $d_x$ and $d_y$ are. Namely, the measuring precision adjustment is feasible through changing the angle of shearing plates, but the choice of measuring precision should consider the distributions of optical vortices as well. Higher measuring accuracy does not mean more distinct intensity ratio points indicating the positions of optical vortices, such as that in the absence of the high intensity ratio point in Fig. \ref{fig4}(a1). As the density of optical vortices increases, the size of the high intensity ratio point becomes smaller. And so does the deviation of the distance between the optical vortex and the high intensity ratio point. That means this measuring method is more suitable for optical vortex measurement with high density. 

In this paper, we used the dual shearing-type Sagnac interferometers for flexible and accurate measurement of the locations of optical vortices in different speckles. The measuring performance was evaluated through comparing the measuring results with that of reference wave based interferometric method. Measuring the positions of optical vortices is simple through calculating the ratio of the interference intensities of the condition with the HWP and that of the condition without the HWP. Comparing the measuring results of two speckles, it is demonstrated that the optimum measuring performance can be achieved through changing the shearing displacements. Due to the upper limit of the precision determined by the inevitable displacement resulting from the optical system, the measuring accuracy only varies in this limited range. However, the upper limit can be expanded through further optimization of the optical system. Besides, if the HWP is replaced by a photorefractive crystal, the common-path structure and the static devices are able to achieve further stability improvement. This interferometers provides an efficient way to observe the distributions of optical vortices in speckles. And its characteristic that higher density of optical vortices leads to higher measuring accuracy makes it suitable for the measurement of close optical vortices. It is expected to be used for observing the evolution behaviors of branch points in adaptive optics and other optical phenomena containing optical vortices. 

\section{Disclosures} The authors declare no conflicts of interest.

\section{reference}

\bibliography{sample}

\begin{thebibliography}{10}
\newcommand{\enquote}[1]{``#1''}

\bibitem{nye1974dislocations}
J.~F. Nye and M.~V. Berry, {\protect\JournalTitle{Proceedings of the Royal
  Society of London. A}} \textbf{336}, 165 (1974).

\bibitem{1998Branch}
D.~L. Fried, {\protect\JournalTitle{Journal of the Optical Society of America
  A}} \textbf{15}, 2759 (1998).

\bibitem{2008Fractality}
K.~O'Holleran, M.~R. Dennis, F.~Flossmann, and M.~J. Padgett,
  {\protect\JournalTitle{Phys. Rev. Lett.}} \textbf{100}, 053902 (2008).

\bibitem{2008Method}
G.~Berkhout and M.~W. Beijersbergen, {\protect\JournalTitle{Physical Review
  Letters}} \textbf{101}, 100801 (2008).

\bibitem{1999Critical}
I.~Freund, {\protect\JournalTitle{Optics Communications}} \textbf{159}, 99
  (1999).

\bibitem{1993Optical}
I.~Freund, N.~Shvartsman, and V.~Freilikher, {\protect\JournalTitle{Optics
  Communications}} \textbf{101}, 247 (1993).

\bibitem{Baranova1983Wave}
N.~B. Baranova, A.~V. Mamaev, N.~F. Pilipetsky, V.~V. Shkunov, and B.~Y.
  Zel'dovich, {\protect\JournalTitle{J. Opt. Soc. Am.}} \textbf{73}, 525
  (1983).

\bibitem{2011Examination}
K.~Patorski and K.~Pokorski, {\protect\JournalTitle{Applied Optics}}
  \textbf{50}, 773 (2011).

\bibitem{2015Detection}
A.~Mobashery, M.~Hajimahmoodzadeh, and H.~R. Fallah,
  {\protect\JournalTitle{Appl Opt}} \textbf{54}, 4732 (2015).

\bibitem{2012Comparison}
K.~Murphy and C.~Dainty, {\protect\JournalTitle{Optics Express}} \textbf{20},
  4988 (2012).

\bibitem{Mingzhou2008Dipole}
M.~Chen and F.~S. Roux, {\protect\JournalTitle{Journal of the Optical Society
  of America A}} \textbf{25}, 1084 (2008).

\bibitem{wu2019wish}
Y.~Wu, M.~K. Sharma, and A.~Veeraraghavan, {\protect\JournalTitle{Light:
  Science \& Applications}} \textbf{8}, 1 (2019).

\bibitem{2010Measuring}
G.~Berkhout and M.~W. Beijersbergen, {\protect\JournalTitle{Optics Express}}
  \textbf{18}, 13836 (2010).

\bibitem{2005wave}
S.~Velghe, J.~Primot, N.~Gu\'{e}rineau, M.~Cohen, and B.~Wattellier,
  {\protect\JournalTitle{Opt. Lett.}} \textbf{30}, 245 (2005).

\bibitem{2010Common}
N.~M. Baba, {\protect\JournalTitle{Optics Letters}} \textbf{35}, 3003 (2010).

\bibitem{goodman2007speckle}
J.~W. Goodman, \emph{Speckle phenomena in optics: theory and applications}
  (Roberts and Company Publishers, 2007).

\bibitem{Fried:98}
D.~L. Fried, {\protect\JournalTitle{J. Opt. Soc. Am. A}} \textbf{15}, 2759
  (1998).

\bibitem{Sanchez:11}
D.~J. Sanchez and D.~W. Oesch, {\protect\JournalTitle{Opt. Express}}
  \textbf{19}, 24596 (2011).

\end{thebibliography}

\bibliographyfullrefs{sample}

\end{document}